# STATE-SPACE BASED MASS EVENT-HISTORY MODEL I: MANY DECISION-MAKING AGENTS WITH ONE TARGET

By Hsieh Fushing,[1] Li Zhu,[1] David I. Shapiro-Ilan,
James F. Campbell and Edwin E. Lewis

*University of California at Davis and USDA-ARS*

A dynamic decision-making system that includes a mass of indistinguishable agents could manifest impressive heterogeneity. This kind of nonhomogeneity is postulated to result from macroscopic behavioral tactics employed by almost all involved agents. A State-Space Based (SSB) mass event-history model is developed here to explore the potential existence of such macroscopic behaviors. By imposing an unobserved internal state-space variable into the system, each individual's event-history is made into a composition of a common state duration and an individual specific time to action. With the common state modeling of the macroscopic behavior, parametric statistical inferences are derived under the current-status data structure and conditional independence assumptions. Identifiability and computation related problems are also addressed. From the dynamic perspectives of system-wise heterogeneity, this SSB mass event-history model is shown to be very distinct from a random effect model via the Principle Component Analysis (PCA) in a numerical experiment. Real data showing the mass invasion by two species of parasitic nematode into two species of host larvae are also analyzed. The analysis results not only are found coherent in the context of the biology of the nematode as a parasite, but also include new quantitative interpretations.

**1. Introduction.** Consider a dynamic decision-making system consisting of many indistinguishable biological organisms, or agents, within a closed environment. Typical examples in biology and ecology include cases of a large fixed number of animals foraging in a common patch, many insect parasites invading a target host, etc. In such dynamic systems, one particularly interesting and also very frequently encountered phenomenon is the

Received September 2007; revised July 2008.

[1]Supported in part by NIH Grant R01AG025218-02.

dramatic heterogeneity among the systems from a sample of presumably identical systems.

The presence of the heterogeneity among systems seems puzzling, as the mass of agents supposedly behave in an unsupervised fashion, and the massiveness of agents, that is, the large number of agents per system, should drive all systems into some sort of homogeneity. On the contrary, great heterogeneities among systems are often observed. One way to untangle this puzzle is to think of the heterogeneity as a manifestation of some self-organized macroscopic behavioral patterns. In this paper we consider a scenario, explained in Section 2, that gives rise to a particular type of self-organized macroscopic behavioral pattern.

Although all agents are nearly identical morphologically and very similar in many key biological constructs that influence the particular decision of interest, the large number of agents could accommodate a distribution of such constructs with a sizeable range. That is, every system could contain individuals that represent both upper and lower extremes that are very far apart from the majority. They are generically called extremists [Crossan, Paterson and Fenton (2007)] or potential leaders [Rands et al. (2003)].

We postulate that the system indeed needs to be ignited by the extremists, after which the remaining majority of followers could quickly perform the event of interest. In this fashion the heterogeneity among systems will be observed due to behavioral differences of extremists in each experimental system. This between-system heterogeneity is then taken as a macroscopic behavioral pattern because early emergence of extremists will give rise to crowding events much sooner than a system having late disclosure on a relevant temporal scale. That is, in general, between-system heterogeneity can be potentially caused by differences in relatively small extreme components within a system that involves a mass of agents. We contend that an accurate depiction of between-system heterogeneity will prove fundamental to understanding the mechanism of a dynamic decision-making system, especially when considering underlying components of extreme nature. In order to successfully extract such information, a new way of modeling this between-system heterogeneity is required, since random effect models *per se* are mechanistically and philosophically less fit to describe the scenario considered above.

Here we address heterogeneity modeling by imposing an internal state-space structure into the dynamic system when vital configuration information about the mass of agents under study is completely missing. Instrumentally all individuals' decisions are correlated because they all share a common system state. For expositional simplicity, we consider a rather simple internal state-space variable that has only two states. Each system sets off with the same first state, and then switches into the other state without recurrence. The first state is termed as the "impermissible state" in which the particular action of interest is unlikely to occur for the majority of agents. Only after this state-space variable changes from the impermissible state into the "permissible state" can a decision leading to an event possibly occur. The duration of the impermissible state is unknown.

Though the dynamics of this state variable are seemingly strict and simple, many biological systems can be described accurately using this construct. One example is the infective juvenile (IJ) of parasitic nematodes invading a host, which motivated our study. The biology of this system is discussed in detail in the next section. We will show that such a state-space variable could physically exist and be reasonably established in an unsupervised fashion.

Based on the above structural assumption of the internal state-space variable, each individual's event-history becomes a composition of a common duration of the impermissible state, shared by all members of the mass of agents contained in the closed system, and an individual specific time to action within the permissible state. We will refer to this composition as the State-Space Based (SSB) mass event-history model. Within this model, the common random impermissible state duration variable is intuitively thought of as the time duration needed for the small group of extremists to work out their pioneering actions which result in the vital signals that are then detected by the rest of the agents. This common random time duration is the source of macroscopic correlation. Furthermore, we assume that given the duration of the impermissible state, random variables of individuals' times to action within the permissible state are independent. This conditional independence construct is the foundation for the statistical inferences proposed and developed here. Its major goal is to decide whether a sample of dynamic decision-making systems really involves a state-space structural heterogeneity.

To achieve our goal, the statistical inference needs to accommodate several inherent data structures. In a study involving a mass of indistinguishable organisms, two difficulties in data collection are often encountered: first, a single individual may be too difficult to be reliably marked and directly observed due to smallness or the lack of proper technology; second, any measurement requires sacrificing the system in one way or the other. In other words, the system has to be terminated at the time when a measurement is taken. The second difficulty severely limits the researchers having only one measurement per system, while the first structure only allows one discrete count at any time point. To accommodate these two data situations, we study only parametric inferences here. The particular parametric version of SSB mass event-history model considered here is a composition of the Weibull model for the impermissible state duration and Logistic model for time to action under the permissible state. Potential extensions of this parametric version are briefly discussed in the Discussion section.

To further enhance our understanding from dynamic perspectives, several distinctions between the SSB mass event-history model and Logistic model with random effects are compiled through a numerical experiment. Via Principle Component Analysis (PCA), great differences in their spectral structures are revealed. Via cross-sectional distribution comparison along the time axis, great differences in variability of mass event-history are also manifested.

This paper is organized as follows. We briefly describe the biology of nematode invasion in Section 2 as the basis of our model structure. The parametric SSB mass event-history model and its corresponding likelihood function developments are discussed in Section 3. Then statistical inferences and the accompanying identifiability and computations are addressed in Section 4. In Section 5 two numerical experiments are conducted: one is to compare the mass event-history model with the random effect model, and the other is a simulation study of the model proposed here. In Section 6 four real data sets of nematode invasion are analyzed. Other related issues, including model extension, are addressed in the discussion section.

**2. Motivating example: Mass of nematode invasion.** Entomopathogenic nematodes (EPN) in the genus Steinernema are soil-dwelling obligate parasites of insects. The infective juvenile stage (IJ) is the only life stage that lives outside the host and its function is to find, assess and finally infect a suitable host [Lewis et al. (2006)]. During the IJ stage, the nematodes are arrested in development with no eating, growth, mating or reproduction; all of these functions take place inside the host. Within hours of entering the host, the IJ nematodes release symbiotic bacteria that kill the host by septicemia and toxemia within a few days. The nematodes develop into adults, mate and produce up to 3 generations inside a single host over the course of about two weeks, when the nutritional value of the host begins to decline and the next cohort of IJs is produced and leaves the host. Missing in this description of the life cycle is the importance of a time frame for infection.

Ten to hundreds of IJs infect a single host, so the first few must lead the invasion and the remaining of the majority follow. This spontaneous emergence of leaders and followers is generally predicted through a dynamic-game of the foraging group [Rands et al. (2003)]. Among the invading herd of IJs, there is risk associated with being the first to invade for two reasons; first, the host can mount an immune response to kill the invading nematode [Li, Cowles and Cowles (2007), Wang and Ganger (1994)] and second, if a single IJ invades and no others follow, mating and successful reproduction cannot take place. There are also risks to invading the host late in the infection generally associated with the declining quality of the nutritive value of the resource. An insect host undergoing infection by entomopathogenic nematodes is a resource with rapidly changing quality and indications thereof.

Thus, the information on which IJs base their decisions (e.g. chemical cues produced by the infection) is dynamic. This is evidenced by the observation that IJs' invasion behavior is not the same toward infected and healthy hosts [Kunkel et al. (2006); Grewal, Lewis and Gangler (1996); Christen et al. (2007)]. To avoid the risks associated with being the first to invade a healthy host, we would hypothesize that all IJs should collectively wait for another to invade, but once the infection is underway, invasion should be permissible for all and proceed rapidly to avoid the risks associated with an old infection.

This emergence of leader-follower behavior among a mass of IJs would collectively result in a lead time before most invasions take place that would be shared by all IJs in the vicinity of the host (or in an individual system). That is, for most IJs, an individuals's event time to invading a host is the lead time, or duration of the impermissible state for invasion, which is a random variable in the nematode example, plus a random time to initiate the action after perceiving signals indicating that the infection has begun. Perception of these signals implies the termination of the lead time and the beginning of the permissible state for invasions. Hence, the collection of all IJs' event times to invasion are indeed related by the sharing of a common lead time variable. Behaviorally speaking, this compositional event time can explain the IJ's "wait-and-see" invasion strategy.

Available technology does not allow measuring an individual IJ's invasion event time because they are less than 1mm in length and live in the soil and are thus too small to be reliably observed or marked. Our measurements of parasite infection patterns were conducted with a large number of IJs (300) and a single host contained in a 15 ml centrifuge tube with 2 ml of sand at the bottom. To estimate the number of IJs that invaded a host, we exposed hosts in this manner for specified durations, then extracted those IJs remaining in the sand by floating them from the sand in water [for detailed description of experimental methods, see Christen et al. (2007) and Lewis et al., unpublished data]. The experimental system is sacrificed at the time of collection; only one measurement per experiment is possible.

A typical data set consists of a collection of counts from the IJs' dichotomous invasion status (invaded or not) from a sample of systems sacrificed at several designated time points with replication. As analyzed later in Section 6, the particular evidence sought here is the significant heterogeneities observed among invasion counts at a time point, especially among counts from experiments in which IJs have a relatively short exposure duration to a host (less than 12 hours). The reason behind analyzing heterogeneity is that values of the lead time variable in different experimental runs should vary to a great degree due to the involvement of a mass of IJs along with the randomly distributed values for this variable. In nature, the mass of IJs could consist of hundreds to thousands of individual IJs. Once an infection has begun, many IJs are likely to follow quickly, resulting in large counts of invaded IJs. This scenario causes the heterogeneity observed in the experiment described. Table 1 below shows a complete real data set and reveals the typical heterogeneity in invading event counts.

To our knowledge, there is no published study that describes such a compositional structure of invading event time of parasites. Confirmation of such a structure of event times would suggest that individual IJs use information of host infection status to make decisions about invasion. Our goal in this paper is to establish a compositional structure that describes the pattern of invasion decisions of infective stage parasites. Such a model will be useful in describing and comparing the decision-making processes of animals when observation of individuals is impractical and the collection of data is destructive to experimental setups. It will also establish a theoretical framework for asking more sophisticated questions about how parasites find, assess and infect their hosts.

We postulate that if the distribution of lead times has a mode not equal to zero, that is, being distinct from the Exponential distribution, then a positive lead time component in IJ invasion event time is established. A mass event-history model is developed for extracting the lead time distribution information in the next section.

**3. State-space based mass event-history model.** Let $\Omega(\omega^M)$ denote a closed system ($\Omega$) containing a mass of $M$ agents ($\omega$'s) and a single target host. A closed system is defined as having no agents transferring in or out of this system. For the system as a whole, macroscopically, $\Omega$ denotes the state-space variable that takes the value "0" for being in the impermissible state and "1" for being in the permissible state at any time point. In the case when each individual agent is equipped with a microscopic time-independent potential or preference phase variable, then each $\omega$ is a Bernoulli random variable taking symbolic value "+" as in the phase of being able to invade the target host, while symbolic value "−" indicates being out of the action phase. It is noted again that only an agent with $\omega = +$ phase could successfully take action upon the target host under the state-space $\Omega = 1$.

TABLE 1
*First 48 hours of exposure: Steinernema feltiae infecting Galleria mellonella. Each number represents the number of IJs that invaded a single host at the indicated exposure durations*

| 2hrs | 4hrs | 6hrs | 8hrs | 10hrs | 15hrs | 20hrs | 24hr | 48hr |
|------|------|------|------|-------|-------|-------|------|------|
| 22   | 2    | 3    | 143  | 177   | 184   | 240   | 215  | 211  |
| 17   | 7    | 70   | 117  | 181   | 206   | 260   | 280  | 260  |
| 12   | 42   | 37   | 163  | 245   | 216   | 261   | 232  | 255  |
| 0    | 39   | 72   | 83   | 194   | 226   | 238   | 198  | 282  |
| 0    | 2    | 86   | 86   | 227   | 163   | 255   | 276  | 295  |
| 0    | 19   | 22   | 63   | 213   | 238   | 259   | 260  | 281  |
| 18   | 0    | 25   | 153  | 162   | 159   | 224   | 254  | 283  |
| 0    | 0    | 51   | 178  | 137   | 155   | 264   | 238  | 260  |
| 0    | 33   | 39   | 86   | 136   | 151   | 247   | 261  | 279  |
| 5    | 0    | 34   | 157  | .     | 158   | 243   | 239  | 286  |

Further, denote $U$ as the random time duration for $\Omega = 0$, which is not directly observable. Since the system's state status is revealed at any time point when the system is sacrificed, $U$, in fact, is observable with its current-status, not its value. Within the system $\Omega(\omega^M)$, hypothetically each agent would give rise to an event time $T$ from time origin to the moment its action is successful upon the target host. Denote the collection of event times as $\{T_m\}_{m=1}^M$. Below we prescribe the SSB mass event-history model on the system $\Omega(\omega^M)$.

**SSB mass event-history model:** Each event time $T_m$ has a compositional form as $T_m = U + S_m$, with $S_m$ being individual specific time to successful action in the permissible state $\Omega = 1$. For an agent in the $\omega_m = -$ phase, $S_m = \infty$. Next, assume the conditional independency for the collection of $\{T_m\}_{m=1}^M$:

**[Conditional-independence]** *Given $U$, $T_m$ is conditionally independent of $T_{m'}$ for all $m \neq m'$.* Under the SSB mass event-history model setting, we further consider parametric distributions for both $U$ and conditional random variable $T_m | U, \omega$ as follows:

A1. Impermissible state duration $U$ is distributed according to the Weibull distribution, denoted by $Weibull(\lambda, \gamma)$;

A2. The conditional survival function of $T_m$ given $U, \omega$ is logistic, that is, for all $t > u$,

$$(3.1) \qquad Pr[T_m > t | U = u, \omega = +] = \frac{1}{1 + e^{\alpha + \beta(t-u)}}.$$

It is known that originally the Weibull distribution was derived as a distribution of extreme events from a system consisting of many components in a reliability context. The logistic regression model assumption for survival times was discussed in Efron (1988) and is practical and typical for count data. The mass of agents has variable potential phases satisfying the following:

A3. If agent's potential phase $\omega$ is a Bernoulli random variable with

$$(3.2) \qquad Pr[\omega = +] = \eta,$$

then this model setting is denoted by the SSB$^+$ mass event-history model. And the SSB mass event-history model that we will use in Section 4s and 5 is essentially a sub-model with $Pr[\omega = +] = \eta \equiv 1$.

Here we develop the likelihood function under the SSB$^+$ mass event-history model setting. Let $N(t)$ denote an event count from a system $\Omega(\omega^M)$ at time $t$. The conditional probabilities for positive count $N(t)(>0)$ are:

$$
\begin{aligned}
Pr&[N(t)|U=u] \\
&= c(t)\left[\frac{\eta e^{\alpha+\beta(t-u)}}{1+e^{\alpha+\beta(t-u)}}\right]^{N(t)}\left[\frac{\eta}{1+e^{\alpha+\beta(t-u)}}+1-\eta\right]^{M-N(t)},
\end{aligned}
\tag{3.3}
$$

where $c(t) = \binom{M}{N(t)}$.

Let $\theta = (\alpha, \beta, \lambda, \gamma, \eta)'$. Under the model assumptions A1, A2 and A3, the amount of likelihood contributed by one event count $N(t)(>0)$ is calculated as

$$
\begin{aligned}
L(\theta|N(t)) &= \int_0^t Pr[N(t)|U=u; \alpha, \beta, \eta]f_U(U=u; \lambda, \gamma)\,du; \\
&= c(t)\int_0^t \left[\frac{\eta e^{\alpha+\beta(t-u)}}{1+e^{\alpha+\beta(t-u)}}\right]^{N(t)} \\
&\qquad \times \left[\frac{\eta}{1+e^{\alpha+\beta(t-u)}}+1-\eta\right]^{M-N(t)}\gamma\lambda^{-\gamma}u^{\gamma-1}e^{-(u/\lambda)^\gamma}\,du.
\end{aligned}
\tag{3.4}
$$

As for zero count $N(t)=0$, the amount of likelihood contributed is equal to

$$
\begin{aligned}
L(\theta&|N(t)(=0)) \\
&= e^{-(t/\lambda)^\gamma} + \int_0^t\left[\frac{\eta}{1+e^{\alpha+\beta(t-u)}}+1-\eta\right]^M\gamma\lambda^{-\gamma}u^{\gamma-1}e^{-(u/\lambda)^\gamma}\,du.
\end{aligned}
\tag{3.5}
$$

Suppose that there is a sample of systems $\{\Omega_{ij}(\omega^M), i=1,\ldots,I; j=1,\ldots,J\}$. Correspondingly, they are sacrificed at time points $\tilde{t} = (t_1,\ldots,t_I)$ with $J$ replications, and result in a sample of counts $\mathcal{N}(\tilde{t}) = \{N_{ij}(t_i)\}$. Then the likelihood function based on data $\mathcal{N}(\tilde{t})$ is computed as

$$
L(\theta|\mathcal{N}(\tilde{t})) = \prod_{i=1}^I \prod_{j=1}^J L(\theta|N_{ij}(t_i)),
\tag{3.6}
$$

where $L(\theta|N_{ij}(t_i))$ is based on (3.4) and (3.5).

Statistical inferences based on $L(\theta|\mathcal{N}(\tilde{t}))$ will be developed in the next section. In advance, it is noted that, due to the large value of $M$, the amount of information for parameters $\alpha, \beta$ and $\eta$ will be significantly larger than that for $\lambda, \gamma$. This feature becomes a characteristic for the SSB/SSB$^+$ mass event-history model setting, since it induces a way to simplify and stabilize maximum likelihood computations involved with numerical integration, as well as high dimensional maximization.

**4. Statistical inferences and computations.** In this section we first address the identifiability and then proceed to discuss the MLE computations for statistical inference of the SSB mass event-history model. For simplicity, we focus on our discussion on the setting of the SSB, not the SSB$^+$, mass event-history model.

4.1. *Identifiability and information content issues.* Since $U$ is not observable in the compositional structure $T_m = U + S_m$, the issues of identifiability and information contents of $\theta = (\alpha, \beta, \lambda, \gamma)$ under the SSB mass event-history model are not entirely obvious and need clarification. The marginal distribution of $N(t)$ computed as $Pr[N(t)|\theta] = L(\theta|N(t))$ is specified for any given time point $t$. This distribution contains the following factor:

$$(4.1) \qquad \Delta(t|\theta) = \int_0^t \left[ \frac{1}{1 + e^{\alpha + \beta(t-u)}} \right]^M \gamma \lambda^{-\gamma} u^{\gamma-1} e^{-(u/\lambda)^\gamma} \, du,$$

which theoretically and practically plays an important role in deciding the amount of information content and sheds light on the identifiability issue as well. Its presence is found through the following two equations:

$$(4.2) \qquad Pr[U < t|\lambda, \gamma] = Pr[N(t) > 0|\theta] + \Delta(t|\theta),$$

$$(4.3) \qquad Pr[N(t) = 0|\theta] = Pr[U \geq t|\lambda, \gamma] + \Delta(t|\theta).$$

It is clear that if $\theta$ and $M$ together in (4.1) make $\Delta(t|\theta)$ very small and ignorable relative to the Weibull survival probability $Pr[U \geq t|\lambda, \gamma]$ at some time points $t_i$, then from (4.2) and (4.3), for all $t_i \in \tilde{t}$, we have $Pr[U \geq t_i|\lambda, \gamma] \approx Pr[N(t_i) = 0|\theta]$. Thus, the parameters $(\lambda, \gamma)$ in the Weibull distribution of $U$ could be extracted with good precision, and so can logistic parameters $(\alpha, \beta)'$. Empirical evidence indicates this is indeed the case when the replicated $N(t_i)$ observed at one time point $t_i$ are highly heterogeneous in the fashion that some systems have rather large numbers of individuals, but some are zeros. This evidence requires that $\alpha$ is not too far from zero in negative value. On the other hand, if $\alpha$ is far below zero, the factor $\Delta(t|\theta)$ can not be too small relative to $Pr[U > t|\lambda, \gamma]$ for most of $t_i$'s. Hence, zero counts should be homogeneously seen among replications. Uniform counts of zero also imply very little information content toward $\theta$.

The above two empirical phenomena, great heterogeneity vs. complete homogeneity, in zero counts constitute evidence borne from the fact that the distribution of $U$ does not mingle with the Logistic distribution, since the latter is in a location-scale family and the former is not. This is the intuition bearing the identifiability issue.

For analytical argument on the identifiability issue, we rewrite the marginal distribution into the following integral form:

$$(4.4) \qquad Pr[N(t) = k|\theta] = \int_0^\infty G_k(t, u|\theta) \gamma \lambda^{-\gamma} u^{\gamma-1} e^{-(u/\lambda)^\gamma} \, du,$$

where

(4.5) $G_k(t, u|\theta) = \begin{cases} \left[\dfrac{\eta e^{\alpha + \beta(t-u)}}{1 + e^{\alpha + \beta(t-u)}}\right]^k \left[\dfrac{1}{1 + e^{\alpha + \beta(t-u)}}\right]^{M-k}, & \text{if } u < t; \\ 0, & \text{if } u \geq t, \end{cases}$

with $k = 0, \ldots, M$.

When $\beta = 0$, the parameters $\alpha$ and $(\lambda, \gamma)'$ are completely separated within the expression of $Pr[N(t) = k|\theta]$. Therefore, it is known that we need at least two time points, that is, $I \geq 2$, to identify of $(\lambda, \gamma)'$. With $I \geq 2$, the above set of integral equations defined through the collection of bounded and linearly independent functions $\{G_k(t, u|\theta)\}_{k=0}^M$ would ensure the identifiability of our SSB mass event-history model. In other words, the equality of marginal distributions $Pr[N(t_i)|\theta] = Pr[N(t_i)|\theta^*]$ should imply the equality of $\theta = \theta^*$.

4.2. *Computations for MLE.* For maximum likelihood estimation (MLE) computations we propose to directly maximize the likelihood function derived in the previous section. However two kinds of computational difficulties face us. First, with the state-space structure imposed with the [*Conditional-Independence*] assumption, the likelihood function, $L(\theta|\mathcal{N}(\bar{t}))$, derived in (3.4)–(3.6) (with $\eta = 1$ for the SSB mass event-history model specifically) involves one-dimensional integration in each of its components. Numerical integration errors resulting from component-wise approximations could sum up to reach a nonignorable level. It is this difficulty that restricts us from employing the EM-algorithm, since the integration error would consequently cause the iteration trajectories in this EM-algorithm to fall into an oscillating phase without converging to a fixed value.

Second, the aforementioned significant difference in information contents between $(\alpha, \beta)$ and $(\lambda, \gamma)$ likely causes the instability of inverting the Hessian matrix within the maximization for the 4-dimensional parameter $\theta$ via the Newton–Raphson method. For these two difficulties, the grid search method is recommended to robustly compute the MLE of $\theta$, denoted as $\hat{\theta}$.

Furthermore, it is interesting and important to make use of unevenness of information contents by carrying out a profiled likelihood type of optimization via grid search. We suggest the following procedure for optimization:

Op1. First, an initial estimate of Weibull parameters $(\lambda, \gamma)'$ could be calculated based on current status data: zero counts of $N(t_i)$ give rise to a right-censored duration of the impermissible state, while positive counts give rise to left-censored data. Denote this initial estimate as $(\hat{\lambda}_0, \hat{\gamma}_0)'$.

Op2. By plugging $(\hat{\lambda}_0, \hat{\gamma}_0)'$ into the full likelihood function $L(\theta|\mathcal{N}(\bar{t}))$, the grid search is performed for an initial estimate of the Logistic parameters $\alpha, \beta$.

Op3. Via the profiled likelihood, we iteratively estimate $(\lambda, \gamma)'$ and $(\alpha, \beta, \eta)'$ once or twice more.

The reason we need only iterate once or twice is that $\alpha, \beta$ can be very well estimated even in the initial estimation.

With the estimate $\hat{\theta}$ that results from the above iterative procedure, we then proceed to compute the observed Fisher information matrix $i(\theta)$ based on the log-likelihood function $l(\theta|\mathcal{N}(\tilde{t})) = \log L(\theta|\mathcal{N}(\tilde{t}))$, as given in the Appendix [Fushing et al. (2008)]. This observed Fisher information matrix $i(\theta)$ could be used for interval estimation purposes.

## 5. Simulation: dynamic differences between the mass event-history and the random effect model.
The random effect model *per se* is the most commonly used methodology to accommodate observed heterogeneity in real data. Often it is used by assuming individual differences following a multivariate Normal distribution as the cause of observed nonhomogeneity. This thinking is not universally applicable, because sometimes, if not most of the time, the observed heterogeneity is inherent and mechanistic. Successfully modeling such mechanistic heterogeneity would advance our scientific understanding and provide new platforms for future new discoveries. Thus, from such perspective, it is of great importance for scientists to be able to discern individual differences from mechanistic heterogeneity, and further, to capture the underlying mechanism properly. In this section we explain this discernment.

One random effect model applicable to our problem setting is the Logistic regression model with a probability of failure given in (3.1), and parameters $\alpha$, $\beta$ are assumed to be random. So the likelihood is calculated as follows:

$$
\begin{aligned}
(5.1) \quad & L^{\mathrm{RE}}(\theta^{\mathrm{RE}}|\mathcal{N}(\tilde{t})) \\
& = \prod_{i=1}^{I} \prod_{j=1}^{J} c(t) \int_{R^2} \left[ \frac{\eta e^{\alpha+\beta(t-u)}}{1+e^{\alpha+\beta(t-u)}} \right]^{N(t)} \\
& \qquad \times \left[ \frac{\eta}{1+e^{\alpha+\beta(t-u)}} + 1 - \eta \right]^{M-N(t)} f(\alpha, \beta|\theta^{\mathrm{RE}}) \, d\alpha \, d\beta.
\end{aligned}
$$

Throughout this section we take $\eta = 1$ for expositional simplicity.

In the first part of this section a computer experiment is devised and performed to characterize the dynamic differences between the SSB mass event-history and the random effect model in generating the mass event count trajectory from individual systems. In the second part another simulation study is conducted to evaluate the performance of profiled likelihood computations under the SSB mass event-history model.

5.1. *Computer experiment for dynamics of SSB mass event-history model.*
The protocol for the computer experiment is as follows:

1. Under the SSB mass event-history model with a chosen vector value of
$\theta_0 = (\alpha_0, \beta_0, \lambda_0, \gamma_0)' = (-3, 0.15, 4, 1.5)'$ and the number of agents $M = 300$, there are 100 independent mass event count trajectories simulated with each replication governed by the following experimental design:

   (a) In the $h$th experiment, $h = 1, \ldots, 100$, one random lead-time $U_h$ is generated from the Weibull distribution with parameter $(\lambda_0, \gamma_0)'$, and 300 random follow-up survival times from the Logistic distribution with parameter $(\alpha_0, \beta_0)'$, denoted as $\{S_{h,m}\}_{m=1}^{300}$.
   (b) For the complete data set $\{T_{h,m} = U_h + S_{h,m}, h = 1, \ldots, 100; m = 1, \ldots, 300\}$, we count the cumulative number, $N_h(\tau)$, of events falling into $[0, \tau + 1)$ for time $\tau = 0, 1, \ldots, 60$ (hrs), and denote the complete trajectory by $N_h = \{N_h(\tau)\}_{\tau=0}^{60}$.

2. To mimic the real data that will be discussed in the next section, only one event count is selected from each of the 100 trajectories. We set $I$ scheduled time points, and randomly divide the 100 trajectories into $I$ groups with the common group size being $J$. With $I = 10$ and $J = 10$, we take the scheduled time points $\{t_i, i = 1, \ldots, I\} = \{2, 4, 6, 8, 10, 12, 24, 36, 48, 60$ (hrs)$\}$. Then within the $j$th group, the cumulative number of events falling into $[0, t_{i+1})$ is recorded as $N_{ij}(t_i)$, for $i = 1, \ldots, I$ and $j = 1, \ldots, J$. Thereby we simulated the data of mass event counts denoted as $\{N_{ij}(t_i)|i = 1, \ldots, I; j = 1, \ldots, J\}$. One simulated sample data set is shown in Table 2 for illustration.

3. We then fit the SSB mass event-history model to the above data $\{N_{ij}(t_i)|i = 1, \ldots, I; j = 1, \ldots, J\}$, and computed the MLE $\hat{\theta}^{SSB}$ of $\theta = (\alpha, \beta, \lambda, \gamma)'$ and the corresponding likelihood value $L(\hat{\theta}|\mathcal{N}(\tilde{t}))$. The MLE $\hat{\theta}^{SSB}$ is computed following the procedure described in Section 4.2.

4. Next we fit the Logistic regression model with Normal random effects on $(\alpha, \beta)'$ assuming

$$\begin{pmatrix} \alpha \\ \beta \end{pmatrix} \sim N\left( \begin{pmatrix} \mu_1 \\ \mu_2 \end{pmatrix}, \begin{pmatrix} \sigma_1^2 & \rho\sigma_1\sigma_2 \\ \rho\sigma_2\sigma_1 & \sigma_2^2 \end{pmatrix} \right).$$

We denote the parameter vector of this random effect model as $\theta^{\mathrm{RE}} = (\mu_1, \mu_2, \rho, \sigma_1, \sigma_2)'$. Compute the MLE $\hat{\theta}^{\mathrm{RE}}$ and the corresponding likelihood value as $L^{\mathrm{RE}}(\hat{\theta}^{\mathrm{RE}}|\mathcal{N}(\tilde{t}))$.

5. With $\hat{\theta}^{\mathrm{RE}}$, 100 random samples of $(\alpha_h, \beta_h)'$ are generated for $h = 1, \ldots, 100$. For each pair of $(\alpha_h, \beta_h)$ and $M = 300$, a complete logistic mass event count trajectory is generated and denoted by $N_h^{\mathrm{RE}}(t)$.

TABLE 2
*A simulated sample: SSB-MEHM with* $(\alpha, \beta, \lambda, \gamma)' = (3.5, 0.15, 4, 1.5)'$

| 2hrs | 4hrs | 6hrs | 8hrs | 12hrs | 16hrs | 20hrs | 30hrs | 45hrs | 60hrs |
|------|------|------|------|-------|-------|-------|-------|-------|-------|
| 15 | 0 | 26 | 30 | 40 | 87 | 104 | 228 | 283 | 294 |
| 18 | 23 | 22 | 29 | 58 | 97 | 120 | 225 | 287 | 296 |
| 22 | 16 | 16 | 24 | 42 | 85 | 132 | 243 | 287 | 299 |
| 0 | 16 | 17 | 19 | 35 | 110 | 119 | 191 | 277 | 300 |
| 0 | 0 | 17 | 44 | 57 | 83 | 115 | 197 | 285 | 300 |
| 0 | 0 | 25 | 33 | 68 | 42 | 119 | 208 | 281 | 295 |
| 0 | 0 | 0 | 39 | 40 | 97 | 132 | 226 | 289 | 299 |
| 0 | 13 | 12 | 38 | 42 | 55 | 130 | 236 | 267 | 300 |
| 27 | 13 | 16 | 29 | 59 | 96 | 67 | 137 | 231 | 300 |
| 21 | 16 | 22 | 20 | 37 | 80 | 118 | 189 | 293 | 299 |

The steps 2, 4 and 5 in the above protocol are used to facilitate a platform for meaningfully comparing two generating dynamics of mass event count trajectory. It is observed that the log-likelihood ratio $\log(\frac{L(\hat{\theta}^{SSB}|\mathcal{N}(\tilde{t}))}{L^{\mathrm{RE}}(\hat{\theta}^{\mathrm{RE}}|\mathcal{N}(\tilde{t}))}) > 60$ in this simulated case. This difference in log-likelihood value is rather significant given that the number of parameters in the random effect model is 5, while the SSB mass event-history model involves only 4 parameters. That is, from either AIC or BIC model selection criteria, mass event count data generated from the SSB mass event-history model would be unlikely mistaken as being generated from the Logistic regression model with random effects.

Further, from the dynamic perspectives, this computer experiment is designed to bring out the following three aspects of characteristic differences between the SSB mass event-history model and Logistic regression model with random effect: first, the longitudinal mean curve of mass event counts; second, the cross-sectional distribution of mass event counts; third, the percentage of total variation explained by principle eigenvectors through the principle component analysis (PCA).

**Longitudinal mean curve:** Two main features of the longitudinal mean curve are informative for dynamics comparison: the event onset and the steepest increment. As shown in Figure 1, especially for the first 10 hour region, the horizontal discrepancy is evident between the "the true mean curve" of $\{N_h\}_{h=1}^{100}$ from the SSB mass event-history model and the mean curve of $\{N_h^{\mathrm{RE}}\}_{h=1}^{100}$ from the Logistic regression model with random effect. This difference implies that the Logistic model tends to predict event onset much earlier than the SSB mass event-history model does. It is also observed that the steepest increment of the mass event count likely occurs ahead of that of the SSB mass event-history model. Ideally, confidence bands should be added onto the mean curves to demonstrate the variation along

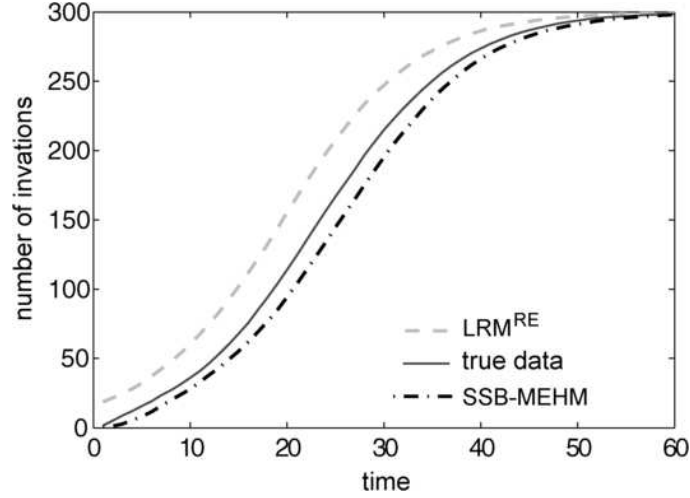

Fɪɢ. 1.  *Comparison of mean curves.*

the temporal axis. We refrain from doing so because the 9 curves contained in the resultant figure become indistinguishable and complicate the visualization of the horizontal difference comparison. As would be understood below and illustrated in Figure 2, the confidence band for the Logistic model with random effect is much narrower than the two related to the SSB mass event-history model.

**Cross-sectional distribution:** Cross-sectional distribution comparison along the temporal axis offers an essential aspect in dynamics comparison. It is particularly informative when two dynamics give rise to very different distribution forms, as seen in Figure 2. We perceive detailed and significant differences in distribution shapes at all the three considered time points. In the 4th hour, there is 60% of cases with no infections in the SSB mass event-history model. In contrast, the Logistic model predicts that all hosts are invaded, and accumulate event counts up to 40. By the 16th hour, the two distributions are centered at different locations with significant different

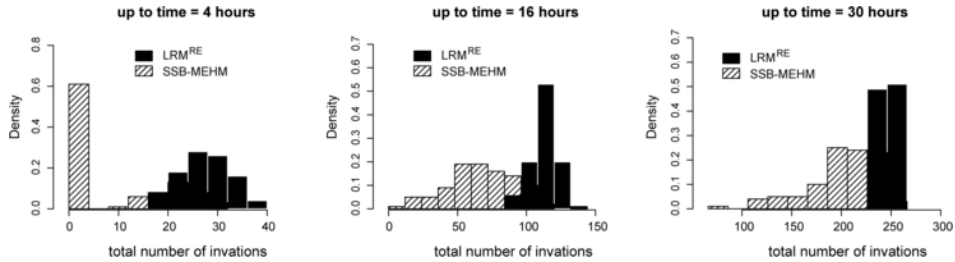

Fɪɢ. 2.  *Frequency comparison of nematodes' invasion counts.*

variations. Further, by the 30th hour, the two distribution forms continue to change in distinct fashions: the Logistic one becomes highly concentrated, while the mass event-history one still drags a heavy and long tail in the left-hand side. These different distribution shapes suggest that the confidence bands for the mean curves in Figure 1 might not be as meaningful as expected when involving only bell shape distributions.

It is worth noting that the heterogeneity revealed in the simulated data $\{N_{ij}(t_i)|i = 1,\ldots,I; j = 1,\ldots,J\}$, or the trajectories $\{N_h\}_{h=1}^{100}$, can be very drastic. In view of the simulated data in Table 2, for example, we see a typical heterogeneity in the simulated vector of event counts at $t_i = 2$ hours: $\{N_{ij}(2)\} = \{15, 18, 22, 0, 0, 0, 0, 0, 27, 21\}$. This kind of heterogeneity is very compatible with that shown in Table 1 of real data. This kind of heterogeneity is indeed observed in three out of four real data sets analyzed in the next section.

**Principle component analysis (PCA):** As the covariance function is a key feature of stochastic processes in general, the temporal covariance matrix provides a characteristic aspect of the dynamic mechanism. By taking each trajectory as a 60-dimensional vector, excluding the 0-hour, the three temporal covariance are computed based on $\{N_h\}_{h=1}^{100}$, $\{N'_h\}_{h=1}^{100}$ and $\{N_h^{RE}\}_{h=1}^{100}$, respectively. Here we use PCA analysis to summarize the $60 \times 60$ temporal covariance matrix for comparing dynamics from the temporal variation perspective.

Three curves of cumulative percentages of total variances explained by the principle components are plotted in Figure 3; the two curves related to the SSB mass event-history models overlap each other. Indeed, Figure 3 provides strong evidence that the Logistic regression model with random effect cannot capture the dynamics governed by the SSB mass event-history model. This conclusion is based on the following evidence. In the Logistic model, the first principle component interprets less than 50% of the total variation of its own invasion trajectories and requires up to the 7th principle component to reach the level achieved by the first principle component of the original 100 simulated invasion trajectories generated via the SSB mass event-history model.

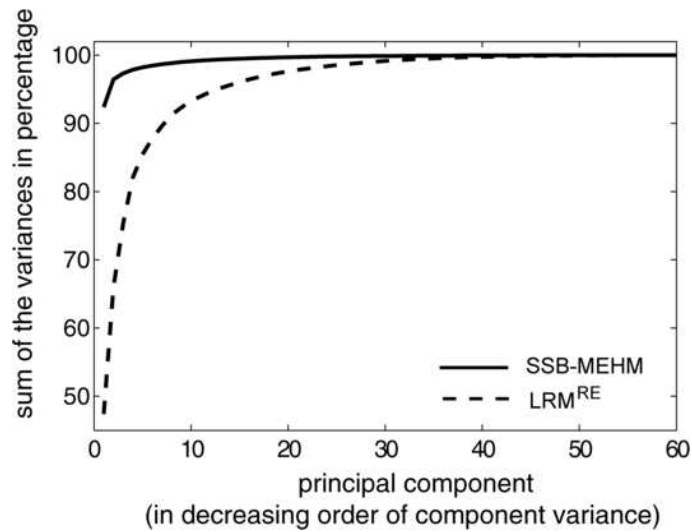

FIG. 3. *Comparison of the cumulative sum of the variances.*

Thus, from the above three critically important aspects, we conclude that the SSB mass event-history model and Logistic model with random effect are very distinct dynamics for generating mass event count trajectories, or time series. Further comparison of these two dynamics through model selection perspective are carried out in real data analysis reported in the next section. Here we reiterate that scientific investigations attempting to accommodate heterogeneity in data should not end at a particular model with random effect. It is essential that models explain mechanisms beyond individual differences.

5.2. *Simulation study for SSB mass event-history model.* A simulation study according to step 3 of the computer experiment in Section 5.1 is performed to evaluate the profiled likelihood computations under the SSB mass event-history model. That is, based on the above simulated data $\{N_{ij}(t_i)|i = 1, \ldots, I; j = 1, \ldots, J\}$, the MLE $\hat{\theta}^{SSB}$ of $\theta = (\alpha, \beta, \lambda, \gamma)'$ is computed via the maximizing profiled likelihood iteration under the likelihood function $L(\hat{\theta}|\mathcal{N}(\tilde{t}))$. The procedure follows Op1–Op3 described in Section 4.2. The results of 300 replications of MLE $\hat{\theta}^{SSB}$ estimations are summarized in Table 3 below. The simulation results confirm that the information content of $\alpha, \beta$ is very different from that of $\lambda, \gamma$, and computations via the profiled likelihood approach work rather well.

TABLE 3
*Summary of parameter estimation based on SSB-MEHM from 300 simulations*

| Parameter | True value | Mean | Standard deviation |
|---|---|---|---|
| $\hat{\lambda}_0$ | 4 | 3.9909 | 0.6235 |
| $\hat{\gamma}_0$ | 1.5 | 1.7394 | 0.6445 |
| $\hat{\alpha}$ | $-3$ | $-3.0050$ | 0.0992 |
| $\hat{\beta}$ | 0.15 | 0.1503 | 0.0051 |
| $\hat{\lambda}$ | 4 | 4.6101 | 0.6343 |
| $\hat{\gamma}$ | 1.5 | 1.6414 | 0.3830 |

**6. Real data analysis and its biological implications.** In this section we analyze four data sets collected from the experimental setup described in Section 2 for two IJs species of nematode (*S. carpocapsae* and *S. feltiae*) vs. two host species (*G. mellonella* and *T. molitor*). In addition to the background provided there, we summarize some biological information regarding interactions between IJs and hosts [for further details, see Lewis et al. (2006)]. This brief summary helps put our data analysis into context.

Once an IJ invades a host, it resumes development which makes the decision to invade the host irreversible. Consequently, the decision directly influences its own fitness plus the environment to be experienced by its offspring. The foraging behaviors of IJs of various species of EPNs differ significantly [Lewis et al. (2006)]: *S. carpocapsae* ambushes hosts by standing on its tail waiting for a passing host; in contrast, *S. feltiae* IJs move through the soil searching for a potential host. The two hosts also differ with respect to their acceptability to each EPN species; *G. mellonella* is the preferred host by both nematode species [Lewis et al. (1996)]. Further, the degrees of interactions between these two IJ species and two host species are not at all uniform: *S. carpocapsae* has poorer performance than *S. feltiae* in *T. molitor*, but has better performance in *G. mellonella*.

Four data sets corresponding to four combinations of IJ and host species with exposure time durations (2, 4, 6, 8, 12, 18, 24, 48)(hrs) are presented here. Interestingly, three of the four data sets contain very heterogeneous mass event count data similar to the data generated from the SBB model in Table 2, the exception being *S. carpocapsae* with the host *G. mellonella*. These four data sets are individually analyzed based on 5 statistical models: (1) Logistic regression model with fixed effect (LRM); (2) Logistic regression model with fixed effect and agent's infectivity phase (LRM$^+$); (3) Logistic regression model with random effect(LRM$^{RE}$); (4) SSB mass event-history model (SSB-MEHM); (5) and SSB$^+$ mass event-history model (SSB$^+$-MEHM).

Consider the Logistic regression model with fixed effect (LRM) as the baseline model. We compare among the five models via the application of Schwarz' (1978) information criterion (BIC). In this application, differences of BIC criterion values between four models and LRM are computed through the formula $(-2)[l(\theta_{\mathrm{model}}|\mathcal{N}(\tilde{t})) - l(\theta_{\mathrm{LRM}}|\mathcal{N}(\tilde{t}))] + (p-2) \times \log(N)$, as reported in Table 4, where $p = \dim(\theta_{\mathrm{model}})$ is the parameter dimension in a "model" among the four models other than LRM ($2 = \dim(\theta_{\mathrm{LRM}})$), and $\mathcal{N}(\tilde{t}) = N$ is the sample size.

Based on the BIC criterion, the SSB$^+$ mass event-history model (SSB$^+$-MEHM) is the model choice for three settings: *S. feltiae* in two hosts species *G. mellonella* and *T. molitor* and *S. carpocapsae* in *T. molitor*. In one setting, *S. carpocapsae* in *G. mellonella*, the BIC selects the Logistic regression model with random effect(LRM$^{RE}$). These results agree with the real data set regarding the presence of heterogeneity in mass event counts like that shown in Table 2. Thus, we can statistically infer that the SSB$^+$-MEHM can best capture the interaction of IJs' behaviors toward host species which give rise to heterogeneous mass event counts.

TABLE 4
*Differences of BIC criterion values from LRM*

| Host | G. mellonella | | T. molitor | |
|---|---|---|---|---|
| Infective Juvenile | S. carpocapsae | S. feltiae | S. carpocapsae | S. feltiae |
| LRM$^+$ | −19.98 | −2727.86 | −1724.24 | −1050.10 |
| LRM$^{RE}$ | −555.96 | −4477.60 | −4418.74 | −2380.30 |
| SSB-MEHM | −471.98 | −4525.74 | −4500.50 | −2422.20 |
| SSB$^+$-MEHM | −459.96 | −4551.60 | −4552.74 | −2446.30 |

Biologically, the results reported in Table 4 support the current understanding of behavioral traits. The majority of individual *S. feltiae* IJs when encountering either species of host are likely adopting a "wait and see" policy. Many risk-averse individuals collectively wait until very few extremists, risk-prone individuals, invade. These features are somehow reflected through having $\gamma$ estimates being significantly different from 1 in Table 5, in which all parameter estimations on all five models are reported. In sharp contrast, the individual infection decisions of *S. carpocapsae* IJs when encountering its favored host, *G. mellonella*, are likely independent from each other. The capability of making a behavioral adjustment for adapting to host differences may not be new, but could be new as a computational outcome: when encountering a less favorable host, such as *T. molitor*, IJs of *S. carpocapsae* adopt the "wait and see" policy similar to *S. feltiae* IJs, and have a very different policy when they encounter a favorable host, such as *G. mellonella*. These results indeed lead to interesting hypothesis for the biology and distribution of EPNs: when nematodes are associated with more susceptible hosts, then distribution should be less aggregated.

TABLE 5
*Estimation comparison*

| Host | | G. mellonella | | T. molitor | |
|---|---|---|---|---|---|
| Infective Juvenile | | S. carpocapsae | S. feltiae | S. carpocapsae | S. feltiae |
| LRM | $\hat{\alpha}$ | −0.4652 | −2.0360 | −1.3312 | −2.5218 |
| | $\hat{\beta}$ | 0.0870 | 0.1694 | 0.0341 | 0.0839 |
| LRM$^+$ | $\hat{\alpha}$ | −0.5353 | −4.9465 | −3.9497 | −3.5473 |
| | $\hat{\beta}$ | 0.1019 | 0.6124 | 0.7279 | 0.2054 |
| | $\hat{\eta}$ | 0.9734 | 0.8477 | 0.4494 | 0.6901 |
| LRM$^{RE}$ | $\hat{\mu_1}$ | −0.4764 | −3.5563 | −2.9760 | −3.6238 |
| | $\hat{\mu_2}$ | 0.0911 | 0.2438 | 0.0649 | 0.1092 |
| | $\hat{\sigma_1}$ | 0.7385 | 0.9970 | 0.9997 | 0.9992 |
| | $\hat{\sigma_2}$ | 0.0451 | 0.0648 | 0.0738 | 0.0123 |
| | $\hat{\rho}$ | −0.8570 | 0.6176 | 0.9966 | 0.8337 |
| SSB-MEHM | $\hat{\alpha}$ | −1.3973 | −3.7067 | −3.0608 | −3.2272 |
| | $\hat{\beta}$ | 0.3223 | 0.5538 | 0.8603 | 0.3821 |
| | $\hat{\lambda}$ | 243.3073 | 212.2950 | 54.9325 | 80.1841 |
| | $\hat{\gamma}$ | 0.9362 | 1.0196 | 2.0094 | 1.8068 |
| SSB$^+$-MEHM | $\hat{\alpha}$ | −1.4321 | −3.8171 | −2.8765 | −3.0832 |
| | $\hat{\beta}$ | 0.3342 | 0.5716 | 1.1390 | 0.5141 |
| | $\hat{\eta}$ | 0.9998 | 0.9659 | 0.7251 | 0.7700 |
| | $\hat{\lambda}$ | 98.0942 | 95.1340 | 60.1455 | 72.6600 |
| | $\hat{\gamma}$ | 1.0011 | 0.7598 | 1.9859 | 1.8790 |

**7. Discussion.** We developed the SSB mass event-history model for modeling potentially self-organized decision-making data obtained from a system constituted of many biological organisms or agents. Such self-organized behaviors in general create macroscopically correlated patterns that underlie a large number of event times within the same system, and render tremendous heterogeneity between replicated systems. This type of manifestation is beyond what general individual-difference based random effect models could accommodate. Our mass event-history models are built with simple internal state-space dynamics for the "wait-and-see" behavioral tactics: the impermissible state represents the behavior of waiting by the extremist or leader to take the first action; the permissible state models the cascade of many followers' decision-making. With this dynamic structure, our mass event-history models shed light on biological and behavioral patterns of decision-making pertaining to many agents sharing a common environment.

From the perspective of statistical merit, our SSB mass event-history models provide a simple and instrumental methodology for accommodating heterogeneity observed among independently and identically designed systems. This capability is distinct from the random effect model *per se*. A random effect model in general maintains a static device for handling heterogeneity stemming from independent, but possibly different individuals within a system. As we demonstrated through simulated as well as real data analysis, the random effect model works well when all involved systems of many agents are rather homogeneous. In contrast, when a system of many agents has the potential to build up system-wise self-organized behaviors, it would be worth modeling the system dynamics by properly capturing the underlying mechanism. Therefore, our mass event-history model is not only an alternative to the random effect model, but an important modeling technique on its own.

From the perspective of dynamic differences, we lay out three temporally oriented aspects in Section 5. Through these aspects, we point out the fundamental differences between the two dynamics governed by the SSB mass event-history model and Logistic regression model with random effect. Although the resultant dynamic differences are informative, we believe that other important and essential perspectives could have been missed in our discussion. Given that this topic of comparing two dynamic systems is not yet well established, research is needed in this direction in statistics.

In the nematode IJ invasion example discussed here, the unobserved state-space variable could be thought of as a physical existence as well as being created in an unsupervised fashion. It may exist as a physiological state that keeps the majority of IJs from making their invading decisions. Only after a few risk-prone extremists or leaders have invaded the host will the rest of the risk-averse majority follow. This type of behavior is indeed seen in finance and many other social sciences. How to properly model the macroscopic correlation resulting from information cascading to the majority is an important issue. Certainly, our internal state-space with impermissible and permissible states would not be sophisticated enough to cope with complexity generated from more intricate decision making systems. Even for IJs, we expect that our simple state-space model structure would become too simple to be suitable when much more informative event-history data than the current-status data are collected. Such a possibility is likely to be seen if advances in experimental and data collection technologies are made possible in the near future.

For current status data collected by sacrificing each experimental system at a time point, the dimensions of model extensions of the SSB mass event-history model could be rather limited. The limitations stem from the compositional and missing data structures involved. The presence of integration in the likelihood function, or marginal probability $Pr[N(t) = k|\theta]$, from time 0 up to several sacrificing time points $t_i$, $i = 1, \ldots, I$, imposes a limit on the number of parameters that are identifiable and estimable. Thus, we need to employ parametric distributions in this setting. Further, as one condition of the SSB mass event-history model, the two compositional distributions involved must not belong to the same family.

However, when the above possibility becomes reality and we could construct a setting where complete individual event-history data are available, the identifiability issue can be alleviated, even while the internal state variable information is still missing. Thus, a modeling extension with one semi-parametric model for time to action in the permissible state and one parametric model for durations under the impermissible state become feasible. Furthermore, the 0-1 internal state-space variable used in the dynamic systems here certainly can be expanded to properly accommodate further complexity of data structure.

## SUPPLEMENTARY MATERIAL

**Score and information** (DOI: [10.1214/08-AOAS189SUPP](#); .pdf). Here we give the gradient and second derivative of the log-likelihood for constructing the score and information, which can be used in numerical estimation of the parameter $\theta = (\alpha, \beta, \lambda, \gamma)'$ in the SSB mass event-history model.

H. FUSHING
L. ZHU
DEPARTMENT OF STATISTICS
UNIVERSITY OF CALIFORNIA
DAVIS, CALIFORNIA 95616
USA

D. I. SHAPIRO-ILAN
USDA-ARS
SOUTHEAST FRUIT
    AND TREE NUT RESEARCH LAB
BYRON, GEORGIA 31008
USA

J. F. Campbell
USDA-ARS
Grain Marketing
    and Production Research Center
Manhattan, Kansas 66502
USA

E. E. Lewis
Department of Entomology
    and Department of Nematology
University of California
Davis, California 95616
USA
E-mail: eelewis@ucdavis.edu